\documentclass[lettersize,journal]{IEEEtran}

\usepackage{cite}
\usepackage{amsmath,amsfonts}
\usepackage[caption=false,font=normalsize,labelfont=sf,textfont=sf]{subfig}
\usepackage{graphicx}
\usepackage{textcomp}
\usepackage{stfloats}
\usepackage{verbatim}
\usepackage{xcolor}

\usepackage{algorithm}
\usepackage{algpseudocode}
\usepackage{booktabs}
\usepackage{multirow}
\usepackage{threeparttable}
\usepackage{tabularx}

\usepackage{xcolor}

\usepackage{balance}
\usepackage{url}
\def\BibTeX{{\rm B\kern-.05em{\sc i\kern-.025em b}\kern-.08em
    T\kern-.1667em\lower.7ex\hbox{E}\kern-.125emX}}
\begin{document}


\title{DynamiQ: Unlocking the Potential of Dynamic Task Allocation in Parallel Fuzzing}

\author{Wenqi~Yan,
        Toby~Murray,
        Benjamin~I.\,P.\ Rubinstein,
        and Van-Thuan~Pham}




\newcommand{\toolname}{DynamiQ}

\maketitle

\begin{abstract}

We present \toolname{}, a full-fledged and optimized successor to AFLTeam that supports dynamic and adaptive parallel fuzzing. Unlike most existing approaches that treat individual seeds as tasks, \toolname{} leverages structural information from the program’s call graph to define tasks and continuously refines task allocation using runtime feedback. This design significantly reduces redundant exploration and enhances fuzzing efficiency at scale. Built on top of the state-of-the-art LibAFL framework, \toolname{} incorporates several practical optimizations in both task allocation and task-aware fuzzing. We evaluate \toolname{} on eighteen targets from Magma and FuzzBench with a total budget of over 40 CPU years. It achieves the highest mean branch coverage on fourteen targets, with average relative gains ranging from 8.37\% to 19.53\% over five baseline fuzzers. Across our experiments, \toolname{} identifies nine previously unknown bugs.

\end{abstract}


\section{Introduction} \label{sec:intro}
\IEEEPARstart{S}{oftware} vulnerabilities remain critical threats to the reliability, security, and integrity of modern software systems. As software complexity and attack surfaces grow, automated methods for systematically uncovering vulnerabilities become increasingly important. Fuzzing, an automated testing technique that generates and executes massive numbers of malformed or unexpected inputs, has emerged as one of the most effective approaches for revealing security bugs~\cite{miller1990empirical}.

Among fuzzing methodologies, coverage-guided greybox fuzzing (CGF) is widely adopted for its effectiveness and efficiency. CGF instruments the target program to collect lightweight execution information—such as branch coverage—to iteratively guide input mutations. Tools such as AFL~\cite{afl}, libFuzzer~\cite{libfuzzer}, and Honggfuzz~\cite{honggfuzz} exemplify CGF’s effectiveness, having uncovered thousands of real-world vulnerabilities in production software~\cite{honggfuzz,ossfuzz}.

Researchers have advanced the effectiveness of fuzzing along two complementary directions. The first focuses on improving fuzzing algorithms through techniques such as smarter seed prioritization~\cite{bohme2016coverage,lemieux2018perffuzz}, taint analysis-guided fuzzing~\cite{rawat2017vuzzer,chen2018angora}, symbolic constraint solving~\cite{stephens2016driller,yun2018qsym}, and structure-aware input generation~\cite{pham2019smart, aschermann2019nautilus}. The second direction, and the primary focus of this paper, seeks to enhance fuzzing efficiency through parallelization. By executing multiple fuzzer instances concurrently across CPU cores or distributed systems, parallel fuzzing aims to scale the testing process and accelerate vulnerability discovery. In principle, this enables faster coverage and bug detection by leveraging modern multi-core hardware. However, in practice, existing parallel fuzzing frameworks often suffer from the \emph{task conflict} problem, where multiple fuzzing instances redundantly explore overlapping program regions due to poor or static task allocation.


Several studies have attempted to address the task conflict problem, with solutions varying based on how they define a fuzzing task. Most existing works consider a task to be a single round of mutation on a seed~\cite{liang2018pafl, wang2021afledge}, focusing on improving seed management, synchronization, and distribution—often through centralized or hierarchical seed management strategies \cite{zhou2020unifuzz,song2019p}. However, as noted by AFLTeam~\cite{pham2021towards}, treating seeds as individual \emph{micro tasks} leads to inefficiencies. Since seeds are largely unrelated, assigning them independently forces fuzzing instances to switch contexts frequently, reducing focus and effectiveness, much like a manager constantly assigning small and unrelated tasks to team members.

To overcome this, AFLTeam~\cite{pham2021towards} proposed defining fuzzing tasks using structural information from the Program Under Test (PUT). Its central hypothesis is that related functions can form useful units of work for parallel fuzzing. It groups these functions by dynamically partitioning the program call graph.

However, we argue that the key assumptions underpinning the dynamic task allocation design of AFLTeam remain insufficiently tested, despite the limited but promising evidence presented in its short paper~\cite{pham2021towards}. Limitations in its design and implementation, together with the limited scope of its evaluation, make those assumptions difficult to assess fully. Its reliance on an older AFL implementation ties the approach to an earlier execution infrastructure. Costly graph partitioning limits frequent task reassignment, while restrictive task handling can prevent workers from pursuing newly emerging coverage opportunities. Section~\ref{sec:background} discusses these limitations in detail.

To address these limitations, we present \toolname{}, a dynamic task allocation framework for parallel fuzzing implemented on LibAFL. Building on the hypothesis introduced by AFLTeam that related functions can form useful fuzzing tasks, \toolname{} continuously refines worker assignments using runtime feedback. Its function scoring model combines remaining uncovered code, recent coverage gains, stagnation, and call graph importance. We incorporate these scores into vertex and edge partitioning to guide task placement and balance partition loads. Selective instrumentation provides coverage feedback from the functions assigned to each worker, while calling context distinguishes executions that reach those functions through different call sequences.

We extensively evaluate \toolname{} on eighteen targets, comprising nine from Magma~\cite{hazimeh2020magma} and nine from FuzzBench~\cite{fuzzbench}, against AFLEdge~\cite{wang2021facilitating}, LibAFL~\cite{fioraldi2022libafl}, Kraken~\cite{zhou2025kraken}, $\mu$FUZZ~\cite{chen2023mufuzz}, and AFLTeam~\cite{pham2021towards}. \toolname{} achieves the highest mean branch coverage on fourteen targets. Its average relative coverage gains range from 8.37\% to 19.53\% across the five baselines.

Our ablation study shows that task partitioning and function scoring contribute to these coverage improvements. Further experiments examine scalability as the core count increases. In terms of bug discovery, \toolname{} finds eight distinct bugs within 24 hours, compared with three to five for each baseline. In an extended fuzzing campaign, \toolname{} also finds nine previously unknown bugs in widely used libraries.

In summary, this paper makes the following contributions:

\begin{itemize}
    \item We design and implement \toolname{}, a full-fledged and practical parallel fuzzing framework that supports dynamic call graph-based task allocation.
    \item We evaluate \toolname{} on eighteen targets with a total budget of over 40 CPU years, measuring branch coverage, the contributions of its allocation and scoring components, scalability, and bug discovery.
    \item We discover 9 previously unknown bugs and vulnerabilities in widely used, well-tested open-source libraries, including sqlite, freetype2, harfbuzz and bloaty.
\end{itemize}

We provide the full reproducibility package of \toolname{}{} at \color{blue} \url{https://github.com/MelbourneFuzzingHub/dynamiq}\color{black}.


\section{Background and Motivation} \label{sec:background}

Parallel fuzzing frameworks typically run multiple fuzzers across CPU cores with shared seed queues, such as in AFL’s monitor-worker mode. While straightforward, this strategy frequently results in redundant exploration, suboptimal utilization of resources, and limited coordination among fuzzing instances. Several proposals have sought to mitigate these issues via centralized coordination or mutation scheduling. However, few use program structure to divide fuzzing tasks.

AFLTeam~\cite{pham2021towards} was one of the first to propose structurally informed parallel fuzzing by partitioning the program call graph and assigning disjoint function subsets to individual fuzzers. To enforce task isolation, AFLTeam precomputes a basic-block level bitmap mask and applies it during seed retention to restrict coverage feedback to relevant partitions. However, its design has several practical and conceptual limitations.

First, AFLTeam characterizes each function using only its covered and total lines of code. Its allocation heuristic can therefore favor larger functions. However, the amount of uncovered code alone does not indicate whether further fuzzing is likely to make progress. In our \texttt{libpng} experiment, 95\% of the functions retained in the call graph showed no further coverage gains after the first 6 hours. Allocating work primarily according to coverage amounts can therefore continue to prioritize regions where exploration has stalled. This limitation is particularly relevant when coverage plateaus and workers must pursue the remaining opportunities.

Second, AFLTeam aggressively prunes the initial call graph generated by static analysis, keeping only functions transitively reachable from the entry point (\texttt{main()}), thereby excluding significant portions of the codebase. For instance, preliminary analysis on \texttt{libxml2} reveals that AFLTeam reduces an original call graph of 2311 functions to merely 357, discarding roughly 85\% of the potential fuzzing space from the outset.

Third, after receiving its assigned partition, each AFLTeam worker filters seeds using precomputed basic block masks derived from the pruned call graph. This filtering identifies inputs relevant to the assigned functions but does not distinguish the calling sequences through which those functions are reached. Although AFLTeam periodically updates the graph through profiling, new coverage in functions outside the current call graph is not represented in these task masks. Inputs that provide new coverage only in those functions can therefore be discarded by the worker, restricting the inputs available for subsequent exploration and graph refinement.

Fourth, AFLTeam adopts Lukes algorithm~\cite{lukes1974efficient} for graph partitioning, a classical method originally designed for tree-structured graphs. This approach scales poorly to real-world call graphs, which are typically cyclic and densely connected. In our measurement, partitioning the \texttt{harfbuzz} call graph, containing approximately 7,000 functions after pruning, takes 8 hours. This cost limits how frequently task assignments can respond to updated runtime information.

Finally, AFLTeam is closely bound to the AFL infrastructure, which limits its ability to take advantage of advanced fuzzing capabilities available in modern frameworks such as LibAFL~\cite{fioraldi2022libafl}. Its evaluation is also limited in scope, providing insufficient empirical evidence to rigorously support the effectiveness of structural task partitioning in parallel fuzzing.

Table~\ref{tab:aflteam-vs-ours} summarizes these limitations and highlights the design improvements introduced by \toolname{}. While the detailed architecture is introduced in later sections, this comparison underscores the motivation for a more adaptive and scalable approach to structural task partitioning.


\begin{table}[t]
\centering
\caption{Comparison of AFLTeam and \toolname{} characteristics.}
\label{tab:aflteam-vs-ours}
\resizebox{\columnwidth}{!}{%
\begin{tabular}{lcc}
\toprule
\textbf{Aspect} & \textbf{AFLTeam} & \textbf{\toolname{}} \\
\midrule
Fuzzing Framework & AFL & LibAFL \\
Graph Partitioning Algorithm & Lukes (vertex only) & Fennel, HDRF (vertex \& edge) \\
Instrumentation Scope & Full Binary & Partition-Aware \\
Function Scoring & Line coverage only & Coverage + centrality + history \\
Seed Retention & Static, pruned-based & Dynamic, inclusive \\
Call Graph Updates & Limited, pruned & Incremental, inclusive \\
New-function seed retention & Discards unseen functions & Retains and incorporates \\
Partitioning Overhead & High ($>$6h for $\sim$6K funcs) & Low (seconds) \\
Coverage Feedback & Edge only & Edge + Call-chain context \\
\bottomrule
\end{tabular}%
}
\end{table}

\section{Approach} \label{sec:approach}

\begin{figure*}[t]
    \centering
    \includegraphics[width=0.92\textwidth]{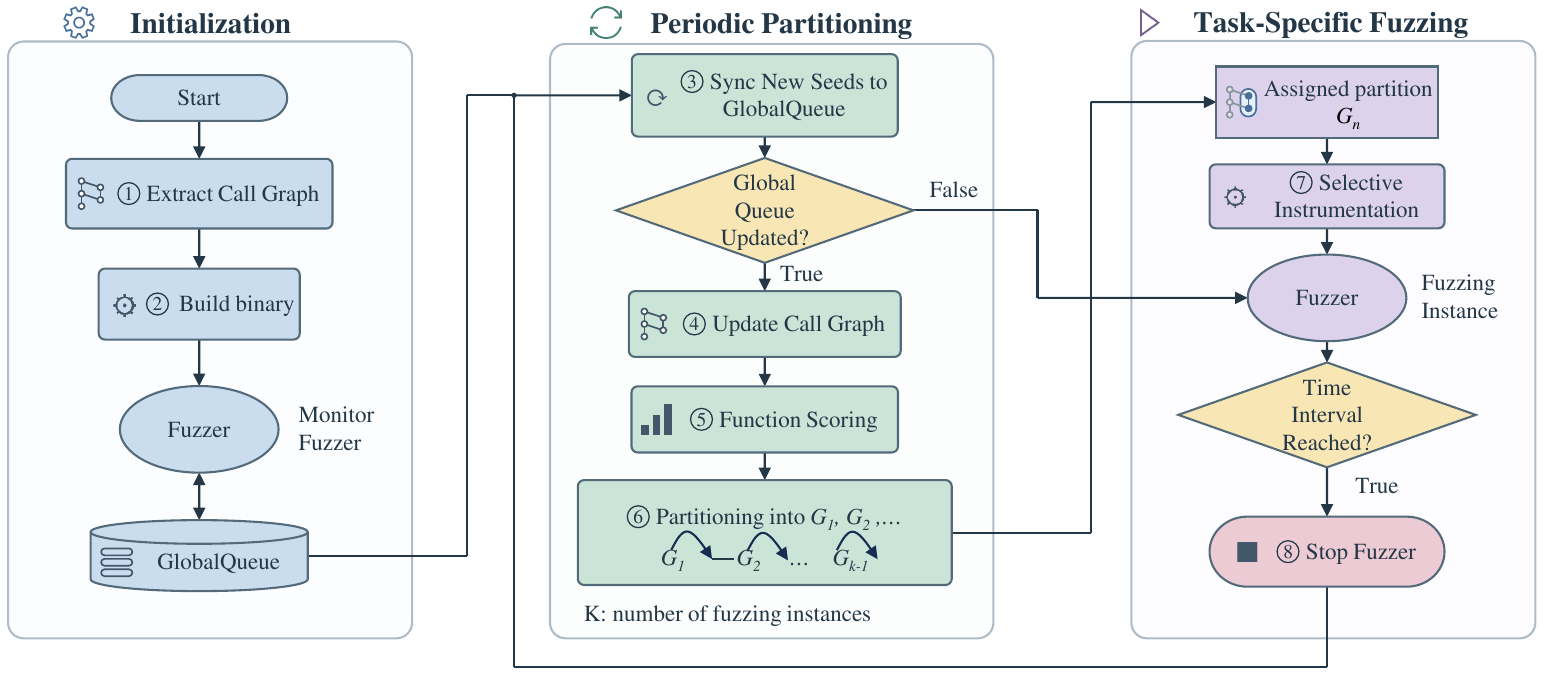}
    \caption{Overview of our dynamic task partitioning framework. The workflow consists of three phases: \textbf{Initialization}, where the call graph is extracted and initial binaries are built; \textbf{Periodic Partitioning}, triggered at some intervals to update the call graph, score functions, and generate partition-specific binaries; and \textbf{Task-Specific Fuzzing}, where each fuzzing instance explores a designated partition.}
    \label{fig:framework-overview}
\end{figure*}

We propose \toolname{}, a dynamic and adaptive parallel fuzzing framework designed to scale efficiently by systematically partitioning the exploration space of the program. Our key insight is that combining runtime feedback with structural properties of the function call graph can effectively guide task allocation, reducing redundancy and accelerating vulnerability discovery. Figure~\ref{fig:framework-overview} outlines our high-level workflow, consisting of Initialization, Periodic Partitioning, and Task-Specific Fuzzing phases.

During \textbf{Initialization}, we construct an initial function call graph using static analysis, compile program binaries with appropriate instrumentation, and launch a monitoring fuzzer. This monitor maintains a global seed queue and periodically synchronizes newly discovered inputs from all fuzzing instances/workers.

In the \textbf{Periodic Partitioning} phase, at regular intervals, the monitor aggregates newly discovered seeds from fuzzing instances, updates the call graph based on runtime coverage traces, scores functions dynamically (Section~\ref{sec:score_partitioning}), and repartitions the program into distinct regions (Section~\ref{sec:partitioning_algorithm}). 

Finally, in the \textbf{Task-Specific Fuzzing} phase (Section~\ref{sec:seed_retention}), each fuzzing instance receives a selectively instrumented binary corresponding to its assigned partition. Coverage feedback from the assigned functions guides which inputs the instance retains for further mutation.

\subsection{Function Scoring} \label{sec:score_partitioning}

To effectively guide dynamic partitioning, we introduce a comprehensive function scoring model that integrates multiple runtime signals: coverage progress, structural importance, and historical exploration difficulty. Prior approaches such as AFLTeam rely primarily on metrics such as raw line coverage, which can favor larger functions. Our scoring function dynamically balances multiple signals using entropy weighting.

We quantify the following metrics for each function \( v \in V \) in the call graph \( G = (V, E) \):

\begin{itemize}
    \item \textbf{Residual Coverage:} \( R(v) = L_{\text{total}}(v) - L_{\text{covered}}^{\text{cur}}(v) \), the number of lines yet to be covered.
    \item \textbf{Recent Coverage Gain:} \( G(v) = L_{\text{covered}}^{\text{cur}}(v) - L_{\text{covered}}^{\text{pre}}(v) \), capturing recent progress in coverage.
    \item \textbf{Exploration difficulty:} \( P(v) = \exp(-0.3 \cdot A(v)) \), where \( A(v) \) is the number of consecutive cycles without new coverage.
    \item \textbf{Structural Importance:} \( C_{\text{katz}}(v) \), the Katz centrality~\cite{katz1953new}, indicating global influence within the call graph.
\end{itemize}

For each function \(v\), the resulting feature vector is
\[
\mathbf{x}(v) = [R(v),\ G(v),\ P(v),\ C_{\text{katz}}(v)].
\]

We normalize each metric across functions using min-max scaling to place the metrics on a common scale~\cite{han2022data}. We then apply entropy weighting to determine their contributions from the distribution of values across functions~\cite{shannon1948mathematical}. Metrics that are constant across functions receive zero weight. Let \(J\) denote the remaining metrics. For each metric \(j \in J\), we compute its entropy as follows.
\[
H_j = - \frac{1}{\log |V|} \sum_{v \in V} p_{vj} \log(p_{vj} + \epsilon), \;
p_{vj} = \frac{x_{vj}}{\sum_{v' \in V} x_{v'j} + \epsilon}
\]
Here, \(x_{vj}\) is the normalized value of metric \(j\) for function \(v\), \(|V|\) is the number of functions, and \(\epsilon\) is a small constant for numerical stability. A lower entropy indicates a less uniform distribution across functions. We normalize \(1-H_j\) over \(J\) to obtain the metric weights \(w_j\).
\[
w_j =
\begin{cases}
\displaystyle\frac{1-H_j}{\sum_{k\in J}(1-H_k)}, & j\in J,\\[6pt]
0, & j\notin J.
\end{cases}
\]

If entropy weighting cannot produce a reliable normalization, we give the remaining metrics equal weights. If no metric distinguishes functions, we assign every function a base score of one.

We recompute the metric weights at each scoring round to reflect the current function measurements rather than use manually fixed coefficients. The function score \(s(v)\) combines the normalized metrics using these weights.
\[
s(v) = \sum_{j} w_j \cdot x_{vj}
\]

The function scores guide partition placement and load balancing at each repartitioning round. Changes in coverage, stagnation, and call graph importance can therefore alter how functions are grouped and assigned to workers.

\subsection{Periodic Partitioning} \label{sec:partitioning_algorithm}

Periodically, the system updates task assignments by repartitioning the call graph according to dynamic function scores. The partitioning algorithm aims to distribute fuzzing potential (function scores) evenly across instances while minimizing calls between partitions. 

\begin{figure}[t]
    \centering
    \includegraphics[width=0.9\linewidth]{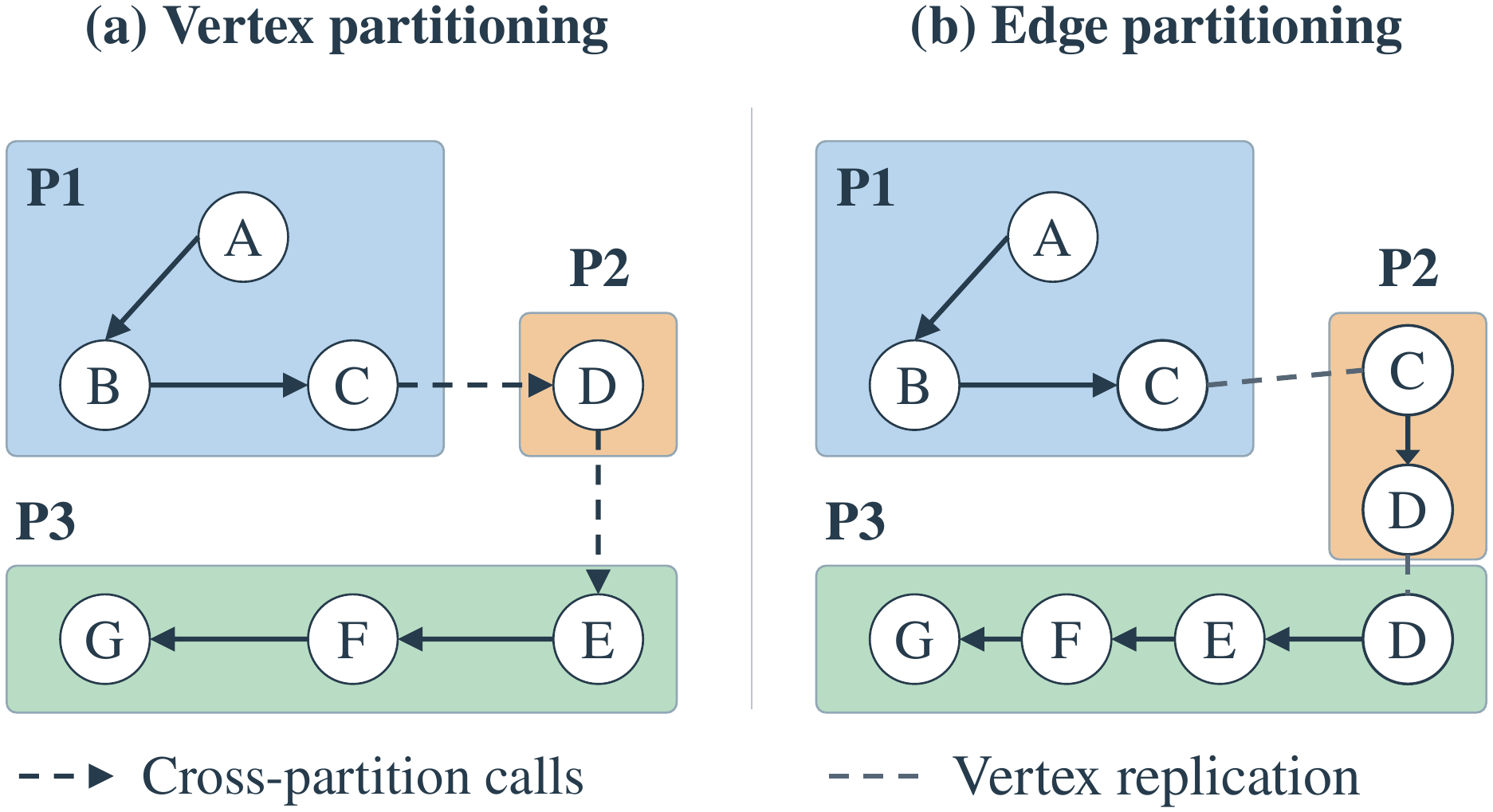}
    \caption{Vertex partitioning (left) assigns each function to one partition. Edge partitioning (right) assigns calls to partitions and allows a function to belong to multiple partitions. Dashed arrows on the left show calls between partitions. Dashed links on the right connect copies of the same function, illustrating vertex replication.}
    \label{fig:partitioning_comparison}
\end{figure}

We explore two general paradigms for partitioning this graph, illustrated in Figure~\ref{fig:partitioning_comparison}. In \textit{vertex partitioning}, each function is assigned to a single partition. The goal is to balance total vertex scores across partitions while minimizing the number of inter-partition calls. This approach encourages each fuzzer to focus on a distinct and self-contained region of the program. In contrast, \textit{edge partitioning} assigns each call edge to a partition, and functions may be replicated across partitions if they are endpoints of edges assigned to multiple partitions. While this may introduce redundancy, the level of vertex replication is typically limited. In practice, we observe that most functions are replicated across only a small number of partitions. 

We use Fennel~\cite{tsourakakis2014fennel} for vertex partitioning and HDRF~\cite{petroni2015hdrf} for edge partitioning. In both implementations, partition load is the sum of the scores of its functions. The scores are updated before each repartitioning round.

Fennel processes functions in descending score order. It favors partitions containing their callees and penalizes partitions with higher load. Each function is assigned to one partition, subject to a load limit. If no partition satisfies the limit, the function is assigned to the partition with the lowest load.

HDRF assigns each call edge to a partition by combining a preference for endpoints already present with a preference for lower load. We use relative function scores in the endpoint preference. An endpoint adds its score to the partition load only when it first enters that partition. A function can therefore appear in several partitions while being counted once within each. These choices use runtime scores to distribute fuzzing work while retaining call relationships.

Prior tree-based algorithms such as Luke’s work well on acyclic structures, but real-world call graphs often contain cycles and dense connections, making such approaches less suitable in practice. Moreover, Luke’s dynamic programming formulation scales poorly on large or highly connected graphs, making it impractical for frequent repartitioning.

\subsection{Task-Specific Fuzzing} \label{sec:seed_retention}

To improve coverage efficiency and reduce redundant exploration, we design each fuzzing instance to specialize in a specific program region. This is achieved through a combination of selective instrumentation and context sensitive call chain tracking.

After graph partitioning, each fuzzing instance receives a task consisting of a subset of functions in the call graph. This assignment focuses its exploration on a specific program region. To enforce this task specialization, we apply selective instrumentation to the functions within its assigned partition. Each instance then starts fuzzing with its selectively instrumented binary. During fuzzing, the instance uses new coverage feedback from its assigned functions to decide which test cases to retain for further mutation. This focuses input selection on the assigned region and reduces redundant exploration across workers.

To further distinguish execution paths within each partition, we incorporate context sensitive call chain tracking. However, recording full call stack contexts throughout the program as in prior work~\cite{wang2019sensitive} is not well suited. Tracking complete calling contexts can produce many distinct coverage states and increase the likelihood of hash collisions in the coverage map of fixed size used by AFL-style fuzzers (typically \(2^{16}\) entries). These collisions can cause different execution contexts to produce the same feedback.

To address these limitations, we constrain the call stack depth \(fn\) individually for each fuzzing instance, using the average shortest path length within its assigned partition as a proxy for structural depth. This connects the amount of context tracked to the structure of the assigned task. Together with selective instrumentation, it distinguishes calling contexts for functions within the task while restricting the depth of context considered.

We implement this context tracking by extending the standard edge coverage mechanism of AFL-style fuzzers with a lightweight call chain hash. Traditional edge coverage computes a bitmap index using the following expression.
\begin{equation}
\texttt{bitmap\_index} = \texttt{cur\_block} \oplus (\texttt{prev\_block} \gg 1).
\end{equation}

We augment this formula with a context aware hash over the call stack within the configured depth.
\begin{equation}
\begin{aligned}
\texttt{bitmap\_index} =\ 
& \texttt{cur\_block} \oplus (\texttt{prev\_block} \gg 1) \\
& \oplus\ \texttt{hash\_callstack}(fn),
\end{aligned}
\end{equation}

\begin{equation}
\texttt{hash\_callstack}(fn) = \bigoplus_{i=1}^{fn} \texttt{hash}(f_i)
\end{equation}

Here, \(f_i\) is the function identifier at stack depth \(i\), and \(\bigoplus\) denotes XOR. This context sensitive feedback distinguishes executions with identical edge transitions but different calling contexts.




\section{Implementation} \label{sec:implementation}

In this section, we describe the implementation details of our dynamic task partitioning framework, \toolname. We start by outlining the detailed workflow in Algorithm~\ref{alg:partitioning}, followed by explanations of the core modules and their interactions, system implementation specifics, and considerations for handling incomplete call graphs.

\subsection{Workflow} \label{sec:workflow}

\begin{algorithm}[t]
\caption{Dynamic Task Partitioning Workflow}
\label{alg:partitioning}
\textbf{Input:} Program Source Code \( P \), Initial Seed Corpus \( S_0 \), Time Interval \( T_{\text{interval}} \), Number of Instances \( K \)
\begin{algorithmic}[1]
\State $G \gets \textsc{ExtractCallGraph}(P)$
\State $(P_{\text{fuzz}}, P_{\text{prof}}, P_{\text{cov}}) \gets \textsc{BuildBinaries}(P)$
\State \textsc{LaunchMonitor}($P_{\text{fuzz}}, S_0$)
\State \textsc{LaunchFuzzers}($P_{\text{fuzz}}, S_0, K{-}1$)
\State $globalQ \gets S_0$;\quad $doneQ \gets \emptyset$
\Repeat
     \If{\textsc{TimeElapsed}() \(\geq T_{\text{interval}}\)}
        \State \textsc{TerminateFuzzers}(\(K - 1\))
        \State $newSeeds \gets globalQ \setminus doneQ$
        \State $G \gets \textsc{UpdateGraph}(P_{\text{prof}}, P_{\text{cov}}, newSeeds, G)$
        \State $doneQ \gets doneQ \cup newSeeds$
        \State $G_{\text{parts}} \gets \textsc{Partition}(G, K{-}1)$
        \State $P_{\text{fuzz}}^{\text{parts}} \gets \textsc{SelectiveInstr}(P, G_{\text{parts}})$
        \State \textsc{LaunchFuzzers}($P_{\text{fuzz}}^{\text{parts}}, globalQ, K{-}1$)
    \EndIf
\Until{\text{TimeoutOrAbort}()}
\end{algorithmic}
\end{algorithm}

We implement our proposed framework, \toolname, as a modular system that follows the dynamic task partitioning workflow shown in Algorithm~\ref{alg:partitioning}. It is composed of three coordinated modules: initialization, periodic partitioning, and task-specific fuzzing.

During initialization (lines~1–5), we extract a static function call graph from the source code and compile three binaries: a profiling binary for tracing dynamic function calls, a coverage binary for collecting line-level or branch-level coverage, and a fuzzing binary for runtime mutation. The monitor fuzzer is launched on one instance to oversee global coordination and manage the global queue, while the remaining \(K{-}1\) instances are initialized as parallel fuzzers using the same initial seed corpus.

At some time intervals (line~7), the monitor fuzzer initiates the task repartitioning cycle. It first terminates all other fuzzers (line~8), identifies newly discovered seeds by comparing the global queue with previously processed inputs (line~9), and refines the call graph using the \textsc{UpdateGraph} procedure (line~10). The graph is then repartitioned into \(K{-}1\) subgraphs (line~12). For each partition, we compile a binary with coverage instrumentation restricted to its assigned functions (line~13). The workers resume fuzzing with these binaries and the updated global seed queue (line~14). This process repeats until timeout or manual termination.

\begin{algorithm}[t]
\caption{\textsc{UpdateGraph}}
\label{alg:updatecallgraph}
\textbf{Input:} Profiling Binary $P_{\text{prof}}$, Coverage Binary $P_{\text{cov}}$, \\
\hspace*{2.7em} New Test Cases $\mathcal{S}_{\text{new}}$, Original Call Graph $G$ \\
\textbf{Output:} Updated Call Graph $G_{\text{updated}}$
\begin{algorithmic}[1]
\State $G' \gets G$, $\mathcal{E} \gets \emptyset$
\State $\mathcal{C} \gets$ \textsc{CurrentCoverage}($G$)
\ForAll{$s \in \mathcal{S}_{\text{new}}$}
    \State $\mathcal{E} \gets \mathcal{E} \cup$ \textsc{RunProfilingBinary}($P_{\text{prof}}, s$)
    \State $c \gets$ \textsc{RunCoverageBinary}($P_{\text{cov}}, s$)
    \State $\mathcal{C} \gets$ \textsc{MergeCoverage}($\mathcal{C}, c$)
\EndFor
\State $G' \gets$ \textsc{CompleteGraph}($G', \mathcal{E}$)
\State $G' \gets$ \textsc{UpdateCoverage}($G', \mathcal{C}$)
\State $G' \gets$ \textsc{ScoreFunctions}($G'$)
\State \Return $G'$
\end{algorithmic}
\end{algorithm}

Algorithm~\ref{alg:updatecallgraph} details the call graph update procedure. We replay newly discovered test cases on the profiling and coverage binaries to collect call relationships and line coverage. After processing these inputs, we update the graph and compare the accumulated coverage with the previous round. We then compute function scores across the updated graph before partitioning.

\subsection{System Implementation}

We implement \toolname{} as a modular system composed of three primary components: initialization, periodic partitioning, and task-specific fuzzing. While the dynamic workflow is outlined in Section~\ref{sec:workflow}, we now describe the engineering details of each component and how they are integrated.

\vspace{0.5em}
\noindent\textbf{Initialization.} We use the LLVM toolchain~\cite{lattner2004llvm} to perform static analysis and extract the initial function call graph from the source code. This graph provides the structural basis for early task assignment. A notable implementation challenge is the incompleteness of the initial static call graph due to indirect function calls, callbacks, and inline assembly. To address this, we extend the \texttt{PCGUARD} instrumentation mechanism in AFL++ to record function-level call edges during execution. 

\vspace{0.5em}
\noindent\textbf{Periodic Partitioning.} The partitioning logic is implemented in Python (approximately 2,128 lines), using NetworkX~\cite{hagberg2008exploring} to represent and manipulate the function call graph. Our implementation supports line, branch, and region coverage. We use line coverage to calculate residual coverage and recent coverage gain because it also provides feedback for functions without branches. For example, a \texttt{sqlite3} static call graph with 3,349 functions contains 605 functions without branch instrumentation that still provide line coverage. Line coverage allows their execution to contribute to function scoring.

To support this dynamic workflow, we restructured and modularized the original AFLTeam codebase, enabling integration of runtime-aware scoring and task reassignment while preserving separation between orchestration, scoring, and instrumentation logic. To further improve robustness, we address limitations in static call graph construction through a secondary mechanism. During the compilation of the profiling binary, we log all functions that contain at least one basic block into a temporary file. This function list captures any potentially reachable functions, regardless of whether their call edges have been observed. Before each partitioning round, we compare this list against the current call graph and conservatively append any missing functions to all partitions. This ensures that such functions are not prematurely excluded, allowing fuzzers to exercise and refine them as execution progresses. Importantly, this step is separate from dynamic call edge collection and serves as a safeguard against under-approximation caused by indirect control flow or assembly.

\vspace{0.5em}
\noindent\textbf{Task-Specific Fuzzing.} Each worker receives a binary compiled with an LLVM instrumentation allowlist for its assigned functions. The worker can execute code throughout the program, but it uses coverage changes within its assigned functions to decide which inputs to retain. Our call-chain-sensitive instrumentation is integrated into this process, allowing fuzzers to distinguish path variants based on calling context and retain semantically meaningful test cases.

\vspace{0.5em}
\noindent\textbf{Orchestration Setup.} To coordinate fuzzing and task synchronization, we adopt a hybrid model. A central AFL++ instance is designated as the monitor fuzzer, chosen for its advanced user interface, crash deduplication, and built-in queue monitoring features. We modify its synchronization logic to rapidly ingest test cases generated by LibAFL fuzzers. The remaining fuzzing instances are implemented in LibAFL~\cite{fioraldi2022libafl}, using forkserver execution. This setup ensures efficient communication and consistent global state updates, while allowing scalable task allocation and principled comparison across configurations. We retain AFL++ as the monitor primarily due to its mature user interface and built-in monitor-mode support (via \texttt{-F}), which simplify integration and enable consistent monitoring across diverse fuzzing instances.

\section{Evaluation} \label{sec:evaluation}

We evaluate \toolname{} to answer the following research questions:

\begin{description}
    \item[\textbf{RQ1:}] How does \toolname{} compare to existing parallel fuzzers in terms of coverage?
    \item[\textbf{RQ2:}] How do task allocation and function scoring affect the coverage of \toolname{}?
    \item[\textbf{RQ3:}] How well does \toolname{} scale with increasing numbers of CPU cores?
    \item[\textbf{RQ4:}] How does \toolname{} compare to existing fuzzers in bug discovery, and can it uncover previously unknown bugs?
\end{description}

\vspace{0.5em}
\noindent\textbf{Benchmarks and seeds.} For the coverage experiments in RQ1 through RQ3, we select all nine targets from Magma~\cite{hazimeh2020magma} and nine from FuzzBench~\cite{fuzzbench}. These libraries cover text rendering, multimedia processing, networking, binary analysis, and data parsing in deployed software. We use the Magma seed corpora for the Magma targets and the OSS-Fuzz seed corpora~\cite{ossfuzz} for the FuzzBench targets. These corpora supply inputs from existing fuzzing workloads, and all fuzzers receive the same initial corpus for each target. Table~\ref{tab:benchmarks} lists the benchmark sources, commit pins, and static call graph sizes.

\begin{table}[t]
\centering
\caption{Benchmark sources and static call graph sizes. Commit pins follow the build Makefile. Nodes and directed edges are reported separately, using the maximum count for each.}
\label{tab:benchmarks}
\small
\setlength{\tabcolsep}{3pt}
\begin{tabular}{llrr}
\toprule
\textbf{Target} & \textbf{Commit} & \textbf{Nodes} & \textbf{Edges} \\
\midrule
\multicolumn{4}{l}{\textit{Magma (nine targets)}} \\
libpng & \texttt{ba980b8} & 447 & 838 \\
libxml2 & \texttt{6645324} & 2,338 & 6,728 \\
sqlite3 & \texttt{4d9384c} & 3,349 & 13,185 \\
php & \texttt{52cc923} & 19,097 & 88,677 \\
openssl & \texttt{bf41baa} & 15,389 & 60,846 \\
poppler & \texttt{1d23101} & 16,636 & 50,133 \\
lua & \texttt{40b76de} & 518 & 1,913 \\
libsndfile & \texttt{68f6c16} & 754 & 1,892 \\
libtiff & \texttt{b40dc6e} & 868 & 2,271 \\
\midrule
\multicolumn{4}{l}{\textit{FuzzBench (nine targets)}} \\
bloaty & \texttt{3f36edb} & 5,863 & 16,519 \\
freetype2 & \texttt{82090e6} & 1,841 & 2,734 \\
harfbuzz & \texttt{a1d9bfe} & 16,489 & 32,541 \\
lcms & \texttt{08f4abb} & 948 & 1,898 \\
libaom & \texttt{3b624af} & 2,114 & 1,365 \\
libjpeg-turbo & \texttt{f29eda6} & 1,213 & 1,241 \\
libpcap & \texttt{bbcbc91} & 654 & 1,572 \\
libxslt & \texttt{7504032} & 1,990 & 5,893 \\
mbedtls & \texttt{b55fd70} & 2,717 & 6,664 \\
\bottomrule
\end{tabular}
\end{table}

\vspace{0.5em}
\noindent\textbf{Baseline fuzzers.} We compare \toolname{} against five parallel fuzzing baselines: (1) \textbf{LibAFL-forkserver}, which adopts LibAFL’s standard multi-core execution model using a forkserver-based executor and Low-Level Message Passing (LLMP). LLMP enables efficient communication between fuzzing instances via shared memory, with a central broker broadcasting updates to connected clients without relying on locks or filesystem sync. It follows the same orchestration setup described in Section~\ref{sec:implementation}, with a centralized AFL++ monitor to provide consistent synchronization and user interface support. This ensures comparability across tools while leveraging existing features for test case exchange and runtime monitoring. \toolname{} is built on LibAFL v0.13.2, and the plain LibAFL baseline uses the same release. (2) \textbf{$\mu$FUZZ}~\cite{chen2023mufuzz}, a parallel fuzzing framework that divides fuzzing into microservices. We patch $\mu$FUZZ to save accepted inputs to disk for replay to measure coverage. (3)~\textbf{AFLTeam}~\cite{pham2021towards}, a structurally informed partitioning framework for parallel fuzzing. (4)~\textbf{AFLEdge}~\cite{wang2021facilitating}, which distributes fuzzing tasks to reduce overlap among parallel instances. (5)~\textbf{Kraken}~\cite{zhou2025kraken}, which allocates fuzzing work according to program behavior.


\begin{table*}[!t]
\centering
\caption{Mean branches covered (Cov.), rounded to integers. Gain is the percentage change of \toolname{} relative to each baseline, with positive values favoring \toolname{}. Bold marks the highest mean per target. NA marks $\mu$FUZZ comparisons affected by OOM failures. Gain and $\hat{A}_{12}$ average equally across available targets.}
\label{tab:branch_coverage}
\small
\setlength{\tabcolsep}{2.5pt}
\begin{tabular}{lrrrrrrrrrrr}
\toprule
\textbf{Target} & \textbf{\toolname{}} & \multicolumn{2}{c}{\textbf{AFLEdge}} & \multicolumn{2}{c}{\textbf{LibAFL}} & \multicolumn{2}{c}{\textbf{Kraken}} & \multicolumn{2}{c}{\textbf{$\mu$FUZZ}} & \multicolumn{2}{c}{\textbf{AFLTeam}} \\
\cmidrule(lr){3-4}\cmidrule(lr){5-6}\cmidrule(lr){7-8}\cmidrule(lr){9-10}\cmidrule(lr){11-12}
 & Cov. & Cov. & Gain & Cov. & Gain & Cov. & Gain & Cov. & Gain & Cov. & Gain \\
\midrule
bloaty & \textbf{6,899} & 6,220 & +10.91\% & 5,530 & +24.74\% & 5,625 & +22.65\% & NA & NA & 5,520 & +24.97\% \\
freetype2 & \textbf{8,750} & 8,237 & +6.23\% & 7,535 & +16.13\% & 8,324 & +5.12\% & 7,786 & +12.39\% & 8,111 & +7.88\% \\
harfbuzz & \textbf{13,635} & 13,282 & +2.65\% & 13,218 & +3.15\% & 12,903 & +5.67\% & NA & NA & 12,887 & +5.80\% \\
lcms & 1,805 & 1,643 & +9.88\% & 1,388 & +30.01\% & \textbf{1,809} & -0.20\% & 1,190 & +51.72\% & 1,229 & +46.84\% \\
libaom & \textbf{11,250} & 9,223 & +21.98\% & 10,817 & +4.00\% & 10,623 & +5.90\% & NA & NA & 8,574 & +31.20\% \\
libjpeg-turbo & \textbf{2,805} & 2,719 & +3.16\% & 2,535 & +10.63\% & 2,704 & +3.71\% & 2,701 & +3.84\% & 2,711 & +3.44\% \\
libpcap & \textbf{2,132} & 1,537 & +38.72\% & 1,842 & +15.78\% & 1,404 & +51.83\% & NA & NA & 1,220 & +74.80\% \\
libpng & \textbf{1,957} & 1,818 & +7.63\% & 1,924 & +1.73\% & 1,950 & +0.35\% & 1,819 & +7.60\% & 1,953 & +0.20\% \\
libxml2 & \textbf{11,062} & 10,778 & +2.64\% & 10,937 & +1.15\% & 10,474 & +5.62\% & NA & NA & 10,828 & +2.16\% \\
libxslt & \textbf{11,842} & 11,029 & +7.37\% & 11,149 & +6.21\% & 11,146 & +6.25\% & 11,001 & +7.65\% & 11,027 & +7.39\% \\
mbedtls & 2,774 & 2,746 & +1.01\% & 2,696 & +2.89\% & \textbf{2,995} & -7.38\% & 2,780 & -0.22\% & 2,722 & +1.91\% \\
sqlite3 & \textbf{33,164} & 30,707 & +8.00\% & 32,244 & +2.85\% & 19,289 & +71.93\% & NA & NA & 26,176 & +26.70\% \\
php & \textbf{18,207} & 17,653 & +3.14\% & 17,527 & +3.88\% & 8,133 & +123.86\% & 16,370 & +11.22\% & 17,786 & +2.37\% \\
openssl & \textbf{9,960} & 8,988 & +10.81\% & 8,568 & +16.25\% & 8,839 & +12.69\% & NA & NA & 8,673 & +14.84\% \\
poppler & 22,888 & 21,137 & +8.29\% & \textbf{23,012} & -0.54\% & 19,726 & +16.03\% & NA & NA & 20,461 & +11.86\% \\
lua & \textbf{3,303} & 3,242 & +1.88\% & 3,261 & +1.28\% & 3,257 & +1.41\% & NA & NA & 3,125 & +5.70\% \\
libsndfile & 2,830 & \textbf{2,854} & -0.85\% & 2,762 & +2.45\% & 2,353 & +20.28\% & NA & NA & 2,827 & +0.11\% \\
libtiff & \textbf{5,734} & 5,347 & +7.24\% & 4,985 & +15.03\% & 5,415 & +5.91\% & 4,811 & +19.19\% & 5,234 & +9.57\% \\
\midrule
\textbf{Mean gain} & & \multicolumn{2}{c}{+8.37\%} & \multicolumn{2}{c}{+8.76\%} & \multicolumn{2}{c}{+19.53\%} & \multicolumn{2}{c}{+14.17\%} & \multicolumn{2}{c}{+15.43\%} \\
\textbf{Mean $\hat{A}_{12}$} & & \multicolumn{2}{c}{0.911} & \multicolumn{2}{c}{0.936} & \multicolumn{2}{c}{0.942} & \multicolumn{2}{c}{0.933} & \multicolumn{2}{c}{0.954} \\
\bottomrule
\end{tabular}
\end{table*}

\begin{figure*}[!t]
    \centering
    \includegraphics[width=\textwidth]{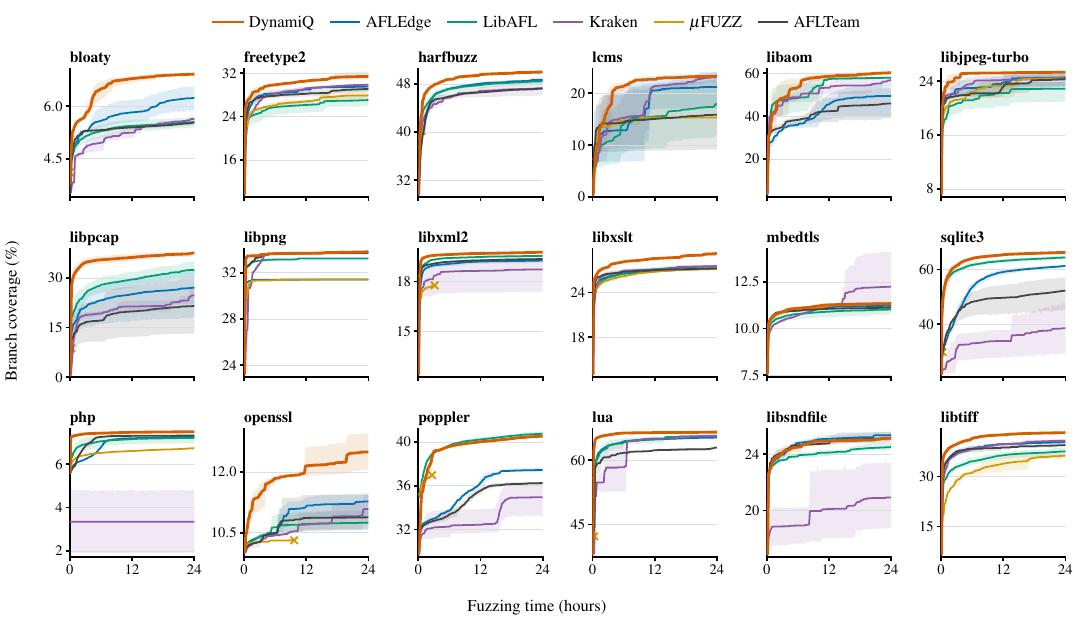}
    \caption{Branch coverage over 24 hours. Lines show means of the plotted runs, and shading gives pointwise 95\% bootstrap confidence intervals. Crosses mark trajectories that end early. Each panel uses its own linear vertical scale.}
    \label{fig:branch_coverage_progression}
\end{figure*}

\vspace{0.5em}
\noindent\textbf{Experimental setup.} We conduct the experiments on machines equipped with AMD EPYC 9474F processors, each configured with 32 vCPUs and 125~GB of RAM, running Ubuntu 22.04. \toolname{} begins with two hours of exploration using the default LibAFL parallel mode with LLMP and forkserver execution. This phase collects coverage and function call information for the first partitioning round. Subsequent repartitioning rounds occur every two hours using updated coverage and profiling data.

At each repartitioning round, \toolname{} replays newly saved inputs on the profiling and coverage binaries to update call relationships and function scores. It partitions the updated graph and recompiles a target binary for each worker with instrumentation restricted to the functions assigned to that worker. Workers then resume fuzzing with the new binaries.

For the \textbf{RQ1} coverage comparison, each fuzzer receives five CPU cores for 24 hours, and we perform ten independent repetitions per target and fuzzer. \toolname{} uses HDRF partitioning. For \textbf{RQ2}, we evaluate the alternative partitioning methods and scoring variants on the same eighteen targets as RQ1 with five cores, 24 hours, and ten repetitions per configuration. For \textbf{RQ3}, we use HDRF at five, ten, and fifteen cores on the same eighteen targets as RQ1. Each configuration runs for 24 hours with ten repetitions and the same initial corpus for each target. For \textbf{RQ4}, we reuse the RQ1 campaigns to compare bug discovery and run \toolname{} once per target for five days using five cores on the latest source commits of the same eighteen targets. The total allocated budget is 356,400 CPU hours (equivalent to over 40 CPU years), counting reused campaigns once and excluding computation outside the campaigns.

\vspace{0.5em}
\noindent\textbf{Performance metrics.} We measure branch coverage in RQ1 through RQ3 and bug discovery in RQ4. To measure coverage, we replay saved inputs on binaries compiled with LLVM coverage instrumentation and collect branch outcomes using \texttt{llvm-cov gcov}. We align input creation timestamps to the campaign clock to obtain coverage over time. The primary comparison uses the same CPU and time budget~\cite{schloegel2024sok}. Execution volume can also affect coverage within this budget~\cite{wang2021riff}, so we also compare coverage at matched execution counts. For bug discovery, we recompile all targets with AddressSanitizer (ASAN).

\subsection{RQ1: Comparison with Existing Parallel Fuzzers}

\subsubsection*{Code Coverage}
Table~\ref{tab:branch_coverage} reports mean branch coverage at 24 hours across eighteen targets. The Gain columns show the percentage by which \toolname{} covers more or fewer branches than each baseline. With HDRF partitioning, \toolname{} achieves the highest mean on fourteen targets. Its mean relative gains are 8.37\% over AFLEdge, 8.76\% over LibAFL, 19.53\% over Kraken, 14.17\% over $\mu$FUZZ, and 15.43\% over AFLTeam. We compute each gain as the arithmetic mean of the relative coverage gains across targets. On ten targets, $\mu$FUZZ terminates before 24 hours because of out of memory (OOM) failures. We mark these entries as NA and compute its mean gain over the remaining eight targets.

The Vargha and Delaney $\hat{A}_{12}$ effect size~\cite{klees2018evaluating} estimates how often a \toolname{} run covers more branches than a baseline run, counting ties as half. Values above 0.5 favor \toolname{}. Table~\ref{tab:branch_coverage} reports its mean across targets.

Parallel workers can revisit the same code when shared inputs guide them toward similar execution paths. \toolname{} assigns workers different functions and uses progress within those functions to select inputs for further mutation. Workers can still reuse seeds found elsewhere, while their assignments direct attention to different parts of the program. The coverage gains are consistent with this division of work, including an improvement over the LibAFL engine on which \toolname{} is built. Figure~\ref{fig:branch_coverage_progression} shows coverage over the campaign.

\subsubsection*{Coverage at Matched Execution Counts}
To assess coverage beyond the effect of execution volume, we also compare \toolname{} with each baseline at the same execution count, following prior work~\cite{yue2020ecofuzz}. For each target and baseline pair, we choose the largest count reached by every run of both tools within 24 hours and compare mean branch coverage at that count. We use the same runs as the coverage over time comparison. The ten $\mu$FUZZ comparisons affected by OOM failures are marked N/A.

\begin{figure}[!ht]
    \centering
    \includegraphics[width=\columnwidth]{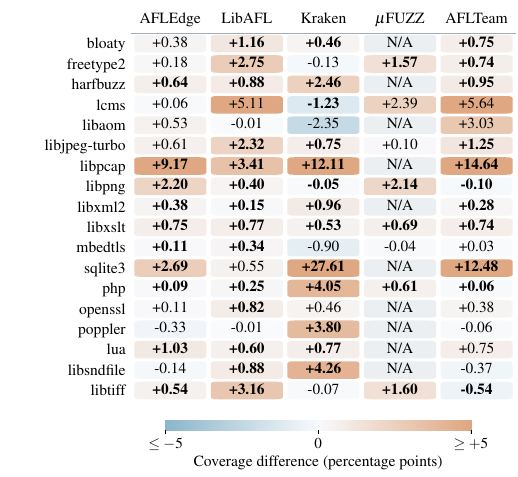}
    \caption{Branch coverage differences at matched execution counts, in percentage points. Positive values favor \toolname{}. Bold marks 95\% bootstrap confidence intervals excluding zero. N/A marks $\mu$FUZZ comparisons affected by OOM failures.}
    \label{fig:matched_execution_summary}
\end{figure}

Figure~\ref{fig:matched_execution_summary} shows higher mean coverage for \toolname{} on sixteen of eighteen targets against AFLEdge, sixteen against LibAFL, twelve against Kraken, and fourteen against AFLTeam. It leads on seven of eight available $\mu$FUZZ targets. These gains at equal execution counts indicate that \toolname{} can reach more branches with the same number of executions, in addition to improving coverage within the 24 hour budget.

\subsection{RQ2: Task Allocation and Function Scoring}\label{sec:rq2}

We examine how task partitioning and function scoring contribute to \toolname{}. We integrate AFLTeam allocation into LibAFL to compare the allocation methods on the same fuzzing engine, and construct variants of \toolname{} that change its partitioner or scoring rule. All configurations share the runtime, graph construction, instrumentation, task filtering, initial corpora, and profiling and repartitioning schedule.

\subsubsection*{Partitioning Algorithm}
We compare three ways of dividing the program among workers: HDRF, Fennel, and random assignment. The random variant distributes functions without considering call relationships or fuzzing potential. It places a main function in each partition and shuffles the remaining functions to balance function counts. This provides a comparison with task division based only on the number of functions.

HDRF achieves the highest mean on fourteen targets (Table~\ref{tab:branch_coverage_algo}). Fennel and random assignment cover 2.88\% and 8.89\% fewer branches, respectively, on average.

Random assignment has the lowest average coverage, indicating that simply giving workers similarly sized sets of functions is insufficient. Functions differ in their remaining coverage opportunities, and arbitrary assignment can separate functions that occur along the same call paths. Both Fennel and HDRF use this information when forming tasks. HDRF further allows a function to belong to multiple partitions when it connects different groups of calls. Its higher average coverage is consistent with retaining these connections while distributing work among workers.

\begin{table}[!t]
\centering
\caption{Mean branches covered by each partitioner. Bold marks the highest mean per target. Change is relative to HDRF.}
\label{tab:branch_coverage_algo}
\small
\setlength{\tabcolsep}{2.5pt}
\begin{tabular}{@{}lrrrrr@{}}
\toprule
\textbf{Target} & \textbf{HDRF} & \multicolumn{2}{c}{Fennel} & \multicolumn{2}{c}{Random} \\
\cmidrule(lr){3-4}\cmidrule(lr){5-6}
 & Cov. & Cov. & Change & Cov. & Change \\
\midrule
bloaty & \textbf{6,899} & 6,612 & -4.16\% & 6,314 & -8.48\% \\
freetype2 & \textbf{8,750} & 8,519 & -2.64\% & 8,172 & -6.61\% \\
harfbuzz & \textbf{13,635} & 13,174 & -3.38\% & 12,046 & -11.65\% \\
lcms & \textbf{1,805} & 1,514 & -16.12\% & 1,028 & -43.05\% \\
libaom & \textbf{11,250} & 10,599 & -5.79\% & 9,936 & -11.68\% \\
libjpeg-turbo & \textbf{2,805} & 2,800 & -0.18\% & 2,641 & -5.85\% \\
libpcap & \textbf{2,132} & 2,026 & -4.97\% & 1,914 & -10.23\% \\
libpng & \textbf{1,957} & 1,952 & -0.26\% & 1,926 & -1.58\% \\
libxml2 & 11,062 & \textbf{11,095} & +0.30\% & 10,382 & -6.15\% \\
libxslt & \textbf{11,842} & 11,572 & -2.28\% & 10,673 & -9.87\% \\
mbedtls & 2,774 & \textbf{2,810} & +1.30\% & 2,683 & -3.28\% \\
sqlite3 & \textbf{33,164} & 32,146 & -3.07\% & 30,862 & -6.94\% \\
php & 18,207 & \textbf{18,235} & +0.15\% & 16,819 & -7.62\% \\
openssl & \textbf{9,960} & 9,583 & -3.79\% & 9,214 & -7.49\% \\
poppler & \textbf{22,888} & 22,361 & -2.30\% & 21,348 & -6.73\% \\
lua & \textbf{3,303} & 3,195 & -3.27\% & 3,028 & -8.33\% \\
libsndfile & 2,830 & 2,814 & -0.57\% & \textbf{2,834} & +0.14\% \\
libtiff & \textbf{5,734} & 5,691 & -0.75\% & 5,471 & -4.59\% \\
\midrule
\textbf{Mean change} &  &  & -2.88\% &  & -8.89\% \\
\bottomrule
\end{tabular}
\end{table}

\subsubsection*{Allocation and Function Scoring}
We compare \toolname{} with three variants to assess its allocation policy, function scorer, and stagnation penalty:
\begin{itemize}
    \item \textbf{AFLTeam (LibAFL)} implements AFLTeam weighting and Lukes partitioning on LibAFL using the shared graph and task feedback.
    \item \textbf{\toolname{} AW} uses AFLTeam weighting in place of the \toolname{} function scorer. The rest of \toolname{} is unchanged.
    \item \textbf{\toolname{} NP} removes the stagnation penalty from the function scorer and recomputes entropy weights over the remaining components.
\end{itemize}

Table~\ref{tab:branch_coverage_variants} compares the variants with \toolname{}. Each variant reduces average coverage, with the largest reduction when both scoring and partitioning are replaced by AFLTeam allocation.

AFLTeam (LibAFL) covers 11.76\% fewer branches even when both allocation methods use the same fuzzing engine and task feedback. This comparison shows a benefit from the allocation policy beyond the choice of engine. AW narrows the change to function scoring. Its 7.56\% reduction indicates that the \toolname{} scorer makes better use of the shared allocation infrastructure by prioritizing functions according to their remaining coverage opportunities, recent progress, and call relationships.

Removing the stagnation penalty reduces mean coverage by 3.12\%. Without this penalty, functions can retain high priority despite repeatedly yielding no new coverage. The lower coverage of NP supports reducing their priority so that other functions receive more attention in subsequent rounds.

Residual coverage and recent coverage gain both derive from coverage measurements, so we examine their association across functions within each scoring round. Across all scoring rounds from ten repetitions of the full \toolname{} configuration on each of the eighteen targets, the Pearson and Spearman correlations are \(0.096 \pm 0.017\) and \(0.087 \pm 0.021\), respectively, reported as mean \(\pm\) standard deviation across rounds. These results indicate a weak average association between the two coverage metrics. The mean metric weights are 29.10\% for residual coverage, 17.91\% for recent coverage gain, 26.45\% for the stagnation penalty, and 26.54\% for structural importance.

\begin{table}[!t]
\centering
\caption{Mean branches covered by the allocation and scoring variants. Bold marks the highest mean per target. Change is relative to full \toolname{}.}
\label{tab:branch_coverage_variants}
\footnotesize
\setlength{\tabcolsep}{1.5pt}
\resizebox{\columnwidth}{!}{
\begin{tabular}{@{}lrrrrrrr@{}}
\toprule
\textbf{Target} & \textbf{\toolname{}} & \multicolumn{2}{c}{\shortstack{AFLTeam\\(LibAFL)}} & \multicolumn{2}{c}{\shortstack{\toolname{}\\AW}} & \multicolumn{2}{c}{\shortstack{\toolname{}\\NP}} \\
\cmidrule(lr){3-4}\cmidrule(lr){5-6}\cmidrule(lr){7-8}
 & Cov. & Cov. & Change & Cov. & Change & Cov. & Change \\
\midrule
bloaty & \textbf{6,899} & 5,493 & -20.38\% & 5,742 & -16.77\% & 6,561 & -4.90\% \\
freetype2 & \textbf{8,750} & 7,788 & -10.99\% & 7,864 & -10.13\% & 8,486 & -3.02\% \\
harfbuzz & \textbf{13,635} & 12,984 & -4.77\% & 13,162 & -3.47\% & 13,348 & -2.10\% \\
lcms & \textbf{1,805} & 1,193 & -33.91\% & 1,271 & -29.58\% & 1,467 & -18.73\% \\
libaom & \textbf{11,250} & 8,864 & -21.21\% & 10,327 & -8.20\% & 10,526 & -6.44\% \\
libjpeg-turbo & \textbf{2,805} & 2,618 & -6.67\% & 2,619 & -6.63\% & 2,789 & -0.57\% \\
libpcap & \textbf{2,132} & 1,328 & -37.71\% & 1,834 & -13.98\% & 2,043 & -4.17\% \\
libpng & \textbf{1,957} & 1,886 & -3.63\% & 1,938 & -0.97\% & 1,952 & -0.26\% \\
libxml2 & 11,062 & 10,793 & -2.43\% & \textbf{11,072} & +0.09\% & 11,043 & -0.17\% \\
libxslt & \textbf{11,842} & 11,048 & -6.70\% & 10,986 & -7.23\% & 11,478 & -3.07\% \\
mbedtls & 2,774 & 2,783 & +0.32\% & 2,718 & -2.02\% & \textbf{2,791} & +0.61\% \\
sqlite3 & \textbf{33,164} & 27,138 & -18.17\% & 31,087 & -6.26\% & 32,204 & -2.89\% \\
php & \textbf{18,207} & 17,562 & -3.54\% & 17,683 & -2.88\% & 18,188 & -0.10\% \\
openssl & \textbf{9,960} & 8,617 & -13.48\% & 8,934 & -10.30\% & 9,579 & -3.83\% \\
poppler & \textbf{22,888} & 20,641 & -9.82\% & 22,349 & -2.35\% & 22,247 & -2.80\% \\
lua & \textbf{3,303} & 3,164 & -4.21\% & 3,171 & -4.00\% & 3,204 & -3.00\% \\
libsndfile & 2,830 & \textbf{2,836} & +0.21\% & 2,788 & -1.48\% & 2,824 & -0.21\% \\
libtiff & \textbf{5,734} & 4,896 & -14.61\% & 5,163 & -9.96\% & 5,708 & -0.45\% \\
\midrule
\textbf{Mean change} &  &  & -11.76\% &  & -7.56\% &  & -3.12\% \\
\bottomrule
\end{tabular}
}
\end{table}

\subsection{RQ3: Scalability with Core Count}

We evaluate \toolname{} with five, ten, and fifteen cores using HDRF at every core count (Table~\ref{tab:branch_coverage_core}). Gains are relative to the five core configuration.

Ten cores increase mean coverage by 3.43\% across targets, and fifteen cores increase it by 6.65\%. Coverage rises on all eighteen targets at both ten and fifteen cores.

\begin{figure}[!t]
    \centering
    \includegraphics[width=\columnwidth]{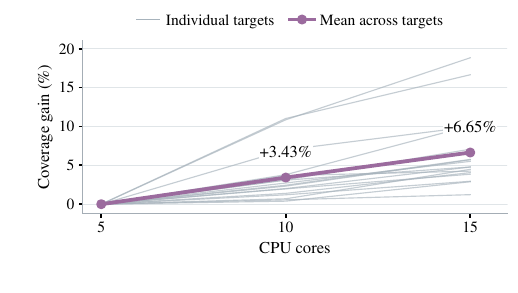}
    \caption{Branch coverage gains relative to five cores. The mean gives each target equal weight.}
    \label{fig:core_scaling_trend}
\end{figure}

The gains show that \toolname{} continues to find additional branches as more workers receive assigned tasks. Figure~\ref{fig:core_scaling_trend} shows a further increase from ten to fifteen cores, rather than a plateau at ten cores. The improvement is modest relative to the additional CPU resources. Tripling the core count yields 6.65\% more coverage.

\begin{table}[!t]
\centering
\caption{Mean branches covered at each core count. Bold marks the highest mean per target. Gain is relative to five cores.}
\label{tab:branch_coverage_core}
\small
\setlength{\tabcolsep}{2.5pt}
\begin{tabular}{@{}lrrrrr@{}}
\toprule
\textbf{Target} & \textbf{5 cores} & \multicolumn{2}{c}{10 cores} & \multicolumn{2}{c}{15 cores} \\
\cmidrule(lr){3-4}\cmidrule(lr){5-6}
 & Cov. & Cov. & Gain & Cov. & Gain \\
\midrule
bloaty & 6,899 & 7,115 & +3.13\% & \textbf{7,387} & +7.07\% \\
freetype2 & 8,750 & 9,361 & +6.98\% & \textbf{9,621} & +9.95\% \\
harfbuzz & 13,635 & 13,909 & +2.01\% & \textbf{14,287} & +4.78\% \\
lcms & 1,805 & 2,001 & +10.86\% & \textbf{2,145} & +18.84\% \\
libaom & 11,250 & 11,598 & +3.09\% & \textbf{11,789} & +4.79\% \\
libjpeg-turbo & 2,805 & 2,816 & +0.39\% & \textbf{2,886} & +2.89\% \\
libpcap & 2,132 & 2,367 & +11.02\% & \textbf{2,487} & +16.65\% \\
libpng & 1,957 & 1,968 & +0.56\% & \textbf{1,981} & +1.23\% \\
libxml2 & 11,062 & 11,280 & +1.97\% & \textbf{11,489} & +3.86\% \\
libxslt & 11,842 & 12,246 & +3.41\% & \textbf{12,644} & +6.77\% \\
mbedtls & 2,774 & 2,876 & +3.68\% & \textbf{2,892} & +4.25\% \\
sqlite3 & 33,164 & 33,626 & +1.39\% & \textbf{34,511} & +4.06\% \\
php & 18,207 & 18,427 & +1.21\% & \textbf{18,744} & +2.95\% \\
openssl & 9,960 & 10,338 & +3.80\% & \textbf{10,965} & +10.09\% \\
poppler & 22,888 & 23,420 & +2.32\% & \textbf{24,196} & +5.71\% \\
lua & 3,303 & 3,393 & +2.72\% & \textbf{3,485} & +5.51\% \\
libsndfile & 2,830 & 2,899 & +2.44\% & \textbf{2,993} & +5.76\% \\
libtiff & 5,734 & 5,774 & +0.70\% & \textbf{5,991} & +4.48\% \\
\midrule
\textbf{Mean gain} &  &  & +3.43\% &  & +6.65\% \\
\bottomrule
\end{tabular}
\end{table}

\subsection{RQ4: Bug Discovery}

We manually triage crashing inputs and remove duplicates based on the top stack frame. Distinct bug counts include both previously unknown and previously reported bugs. A bug is classified as previously unknown (zero day) if it had not been reported upstream when first discovered in our experiments.

\subsubsection*{Comparison with Baseline Fuzzers}
We compare bug discovery using the same 24 hour campaigns with five cores per fuzzer as in RQ1. Table~\ref{tab:bug_discovery_summary} reports 8 distinct bugs for \toolname{}, 5 for LibAFL, 4 each for AFLEdge and AFLTeam, and 3 each for Kraken and $\mu$FUZZ. We also report timeouts and out of memory (OOM) events. An input is classified as a timeout if it runs for more than 20 seconds and as an OOM event if it consumes more than 10~GiB of memory.

\toolname{} finds more distinct bugs than the baselines within this budget. Three of its four bugs in \texttt{bloaty} were previously unknown.

\begin{table}[t]
\centering
\caption{Bug discovery and stability issues per target. Entries show distinct bugs (B) and timeouts or OOM events (T) during the RQ1 campaigns using five cores. The thresholds are 20 seconds and 10~GiB of memory.}
\label{tab:bug_discovery_summary}
\footnotesize
\setlength{\tabcolsep}{2pt}
\resizebox{\columnwidth}{!}{
\begin{tabular}{lcccccc}
\toprule
\textbf{Target} & \textbf{\toolname{}} & \textbf{AFLEdge} & \textbf{LibAFL} & \textbf{Kraken} & \textbf{$\mu$FUZZ} & \textbf{AFLTeam} \\
\midrule
libxml2       & \textbf{(4, 0)} & (2, 0) & (3, 0) & (2, 0) & (2, 0) & (2, 0) \\
sqlite        & \textbf{(0, 12)} & (0, 8) & (0, 3) & (0, 1) & (0, 0) & (0, 7) \\
bloaty        & \textbf{(4, 0)} & (2, 0) & (2, 0) & (1, 0) & (1, 0) & (2, 0) \\
libpcap       & \textbf{(0, 12)} & (0, 4) & (0, 5) & (0, 0) & (0, 0) & (0, 3) \\
harfbuzz      & \textbf{(0, 15)} & (0, 6) & (0, 9) & (0, 0) & (0, 0) & (0, 7) \\
\midrule
\textbf{Total} & \textbf{(8, 39)} & (4, 18) & (5, 17) & (3, 1) & (3, 0) & (4, 17) \\
\bottomrule
\end{tabular}
}
\end{table}


\subsubsection*{Extended Fuzzing Campaign}

To assess the practical use of \toolname{}, we run a five day fuzzing campaign on the same eighteen targets as RQ1 using five CPU cores and the latest source commits available at the start of the campaign.

The extended campaign finds twelve distinct bugs, including six additional previously unknown bugs (Table~\ref{tab:campaign_bug_summary}). These six discoveries comprise one assertion failure, one division by zero, one infinite loop, two null pointer dereferences, and one write outside a buffer, bringing the total to nine previously unknown bugs.

All targets are actively fuzzed by OSS-Fuzz with the same fuzz drivers. \toolname{}’s ability to uncover new bugs highlights the effectiveness of its dynamic, structure-aware partitioning, which reduces redundancy and improves depth of coverage beyond traditional parallel fuzzing.

\begin{table}[t]
\centering
\caption{Distinct bugs found in the extended campaign, classified by their status at first discovery in our experiments.}
\label{tab:campaign_bug_summary}
\small
\renewcommand{\arraystretch}{1.1}
\begin{tabularx}{\linewidth}{lcc>{\raggedright\arraybackslash}X}
\toprule
\textbf{Target} & \textbf{Reported} & \shortstack{\textbf{Previously}\\\textbf{unknown}} & \textbf{Bug types} \\
\midrule
sqlite     & 0 & 1 & Assertion failure \mbox{(CWE-617)} \\
freetype2  & 2 & 2 & Division by zero \mbox{(CWE-369)} \\
           &   &   & Infinite loop \mbox{(CWE-835)} \\
harfbuzz & 0 & 3 & Null pointer dereference \mbox{(CWE-476)} \\
         &   &   & Write outside a buffer \mbox{(CWE-787)} \\
bloaty     & 1 & 3$^{\dagger}$ & Null pointer dereference \mbox{(CWE-476)} \\
\midrule
\textbf{Total} & \textbf{3} & \textbf{9} & \\
\bottomrule
\end{tabularx}
\par\smallskip
{\footnotesize\raggedright $^{\dagger}$The three bloaty bugs were first discovered during the 24 hour comparison and rediscovered during extended fuzzing.\par}
\end{table}

\section{Discussion} \label{sec:discussion}

While \toolname{} improves fuzzing efficiency and scalability, it has limitations that suggest future enhancements—specifically, integrating directed fuzzing into partitioned workflows and enabling adaptive control over repartitioning intervals.

\paragraph{Seed Sharing Across Partitions}
\toolname{} uses coverage within assigned functions to guide input retention. An input that reveals new coverage only outside a worker partition may therefore not be retained or shared. This creates a tradeoff between worker specialization and retaining inputs that could benefit other workers. The coverage gains in RQ1 show that the overall design remains effective on the evaluated targets. A possible extension is to let the monitor sample discarded worker inputs and retain those that reveal new coverage elsewhere in the program.

\paragraph{Directed Fuzzing Integration}
Selective instrumentation and context sensitive call chain tracking shape which inputs each worker retains for further mutation within its assigned partition. Directed fuzzing could provide additional guidance by favoring inputs whose executions approach uncovered functions in that partition. Distance information from techniques such as AFLGo~\cite{bohme2017directed} and Hawkeye~\cite{chen2018hawkeye} could help workers reach functions guarded by specific input conditions. Combining this guidance with task allocation could help workers pursue coverage opportunities within their assigned regions.

\paragraph{Adaptive Partitioning Frequency}
Our evaluation uses static task repartitioning every two hours for simplicity, but this ignores runtime signals that could prompt smarter adjustments. Fixed intervals may waste resources on stalled tasks or disrupt productive ones.

A better approach is dynamic repartitioning based on signals like stagnant coverage, low novelty rates, or workload imbalance. For example, if an instance stops finding new paths, its partition can be reassigned; if another shows high discovery, it can be given more resources. This can be achieved using coverage deltas or adaptive timers informed by online metrics.

\section{Related Work} \label{sec:related}
\textbf{Coverage-guided fuzzing.} Coverage-guided fuzzing is a leading approach in vulnerability discovery, using code coverage feedback to adaptively generate test inputs. AFL~\cite{afl} popularized this method with lightweight instrumentation and input mutation. libFuzzer~\cite{libfuzzer}, part of LLVM, emphasizes fast in-process fuzzing and integration with sanitizers. Honggfuzz~\cite{honggfuzz} extends this by incorporating additional feedback signals, such as hardware performance counters. VUzzer~\cite{rawat2017vuzzer} further advances the field with application-aware fuzzing, leveraging static and dynamic analysis to guide deeper and more targeted mutations based on control- and data-flow features.

\textbf{Parallel fuzzing.} As fuzzing evolves, there is growing interest in scaling it using multi-core and distributed systems through parallel fuzzing. Tools like P-Fuzz~\cite{song2019p} and UniFuzz~\cite{zhou2020unifuzz} use centralized databases to manage seeds and avoid task duplication. PAFL~\cite{liang2018pafl} synchronizes guidance data and divides the coverage bitmap into regions assigned to different instances. glibFuzzer~\cite{gu2022group} groups the global corpus and selects inputs for workers using seed size, coverage information, and differences between groups. AFLEdge~\cite{wang2021facilitating} treats each full mutation cycle on a unique seed as a task and uses edge coverage for dynamic task generation. \toolname{} uses call relationships and function scores to form worker tasks and updates their assignments as fuzzing progresses. AFLTeam~\cite{pham2021towards} leverages attributed call graphs and graph partitioning to guide task allocation. Mufuzz~\cite{chen2023mufuzz} adopts a microservice model, improving scalability through dynamic resource allocation. Dodrio~\cite{liang2024dodrio} introduces redundancy-free scheduling with a dual bitmap system, enhancing parallel taint analysis and task uniqueness. Kraken~\cite{zhou2025kraken} uses runtime feedback to adjust the number of active workers and the input selection strategy.

\textbf{Collaborative fuzzing.} Collaborative fuzzing, or ensemble fuzzing, improves vulnerability detection by combining multiple fuzzers, each with unique strengths. By sharing seeds and test cases, this approach can achieve greater code coverage than any individual fuzzer. EnFuzz~\cite{chen2019enfuzz} first demonstrated this benefit by synchronizing diverse fuzzers to boost overall performance. CollabFuzz~\cite{osterlund2021collabfuzz} extended this with centralized scheduling to reduce redundancy and optimize input distribution. AutoFz~\cite{fu2023autofz} further advanced the idea by automating fuzzer selection and coordination based on the target software and vulnerability characteristics, increasing adaptability across diverse environments.
\section{Conclusion} \label{sec:conclusion}
This paper presents \toolname{}, a practical framework that enables dynamic task allocation in parallel fuzzing. While prior work highlighted the benefits of structure-aware task definitions, existing solutions lacked the scalability, precision, and adaptability needed in practice. Built on LibAFL, \toolname{} combines call graph-based task partitioning, runtime feedback-driven refinement, and task-aware fuzzing strategies. Our evaluation on eighteen Magma and FuzzBench targets shows average relative branch coverage gains of 8.37\% to 19.53\% over five baseline fuzzers. \toolname{} also identifies nine previously unknown bugs.

By addressing key limitations of prior work and showing the effectiveness of dynamic task allocation at scale, \toolname{} advances the state of the art in parallel fuzzing and provides a solid foundation for future research on adaptive and efficient fuzzing strategies.

\section{Acknowledgement}
This work was partially supported by the Australian Research Council’s
Discovery Early Career Researcher Award (DECRA), project number
DE230100473.

\bibliographystyle{IEEEtran}
\balance{}
\bibliography{references}

\end{document}